\documentstyle[preprint,aps,floats,epsfig]{revtex}

\newcommand{\be}{\begin{equation}}
\newcommand{\ee}{\end{equation}}
\newcommand{\bea}{\begin{eqnarray}}
\newcommand{\eea}{\end{eqnarray}}
\newcommand{\bml}{\begin{mathletters}}
\newcommand{\eml}{\end{mathletters}}

\begin{document}

\titlepage

\preprint{gr-qc/0108071}

\bigskip

\draft

\tighten

\title{Dimensional Dependence of Black Hole Formation\\
in Self-Similar Collapse of Scalar Field}

\author{
{Hyeong-Chan Kim}$^a$
         \footnote{E-mail: \texttt{hckim$@$phya.yonsei.ac.kr}},
{Sei-Hoon Moon}$^b$
         \footnote{E-mail: \texttt{jeollo$@$phya.yonsei.ac.kr}}
and {Jae Hyung Yee}$^{a,b}$
\footnote{E-mail: \texttt{jhyee$@$phya.yonsei.ac.kr}}}

\address{$^a$Institute of Basic Science, Yonsei University,
Seoul 120-749, Korea\\
$^b$Institute of Physics and Applied Physics, Yonsei University,
Seoul 120-749, Korea }

\date{\today}

\setlength{\footnotesep}{0.5\footnotesep}

\maketitle

\begin{abstract}
We study classical and quantum self-similar collapses of a massless
scalar field in higher dimensions, and examine how the increase in
the number of dimensions affects gravitational collapse and black
hole formation. Higher dimensions seem to favor formation of black
hole rather than other final states, in that the initial data space
for black hole formation enlarges as dimension increases. On the other
hand, the quantum gravity effect on the collapse lessens as dimension
increases. We also discuss the gravitational collapse in a brane world
with large but compact extra dimensions.
\end{abstract}

\vspace{10mm}

\vspace{5mm}

\indent\indent\hspace{4mm} {Keywords:} gravitational collapse,
self-similarity, black hole, quantum

\indent\indent\hspace{4mm} gravity, higher dimension, large extra
dimensions.


\newpage

\section{Introduction}

Since the Choptuik's discovery \cite{choptuik} of critical phenomena
in gravitational collapse of a scalar field, there have been much
attention to the gravitational collapse just at the threshold of
black hole formation. Similar phenomena to Choptuik's results
were quickly found in other systems numerically \cite{Abra}-\cite{olab}
and analytically \cite{roberts}-\cite{Gundlach1995}
too,\footnote{ See the excellent review by Gundlach \cite{Review},
and more references there in.} suggesting that they were limited
neither to scalar field matter nor to spherical symmetry.
There are two possible final outcomes of gravitational collapses,
distinguished by whether or not a black hole is formed in the
course of evolution. Which end state is realized depends on
initial data specified by a control parameter $p$ that
characterizes the gravitational interaction strength in the ensuing
evolution. For $p<p^*$ the gravitational field is too weak to form
a black hole, while for $p>p^*$ a black hole is produced.
The critical solution with $p=p^*$ just at the threshold of
black hole formation acts as an intermediate attractor for nearby
initial conditions and has an additional symmetry: discrete or
continuous self-similarity.

While critical gravitational collapse has been originally studied
in four spacetime dimensions, there have been several attempts to
study it in higher-dimensional spacetime \cite{soda}-\cite{Villas}.
As most attempts (including string theory) to unify gravity with
other forces require additional spatial dimensions, it may be
meaningful to consider higher dimensional collapse problem.
Although the extra dimensions are compactified to a small size and
usually unobservable at low energy, the extra dimensions may be
opened in the large energy limit of the Planck scale, which is
relevant as we approach singularity in gravitational collapse.

Moreover, in recent years there have been explosive interests in
the large extra dimension scenario \cite{ADD}, in which the size
of extra dimensions can be as big as sub-millimeter and the
fundamental scale of the higher-dimensional gravity as low as a
TeV range. An exciting consequence of this scenario is the possibility
of production of black hole at realistic future colliders \cite{ADDBH}.
The Schwarzschild radius of so produced black hole will be much less
than the size of extra dimensions because the mass of the black hole
is typically of the order of the higher-dimensional Planck scale of TeV.
Therefore, in this scheme the gravitational collapse will be essentially
higher-dimensional and quantum gravitational. We can also expect the
formation of black holes with much larger mass than the fundamental
mass scale, but with smaller Schwarzschild radius than the size of
extra dimensions. Then, black holes can be well treated as general
relativistic objects in higher dimensions. Such a black hole may be
formed from a macroscopic initial distribution of matter prepared
in four-dimensional world, but whether black hole are formed
from the given initial data depends on the number of extra dimensions,
because the final stage of collapse will be entirely higher-dimensional.
Therefore, we expect that one can determine the number of large
extra dimensions by investigating the gravitational collapse and the
dimensional dependence of black hole formation, at least, in principle.
In this sense it will be valuable to study the gravitational collapse
in higher dimensions and the dimensional dependence of black hole
formation.

In the next section, we shall start with a brief summary of Frolov's
works \cite{Frolov}, which treat the higher-dimensional scalar field
collapse with restriction to the spherical symmetry and self-similarity
and examine an analytic solution. We then examine the role of
dimensionality of spacetime in the gravitational collapse and
black hole formation. Higher dimensions seem to favor black hole in
that, as dimension increases, the critical value of the control
parameter decreases and so the initial data space for black hole
formation enlarges, while the size of the apparent horizon increases
near the critical collapse.

We will also consider, in Section \ref{III}, the quantum gravity effects
on the collapse of the scalar field by extending the work of
Ref.\cite{tomimatsu,BKKSY} to higher dimensions. Since the mass of the
black hole near the critical collapse can be arbitrarily small, the
quantum gravity effects may play a significant role in such collapse.
We also examine the role of spacetime dimensionality in such quantum
collapse. The quantum gravity effects on the collapse decrease as the
spacetime dimension increases. Finally, in the last section, we discuss
the gravitational collapse in the brane world with large extra
dimensions.

\section{Dimensional dependence of gravitational collapse}\label{II}

We begin this section with a short review of Frolov's solutions of
Ref.\cite{Frolov} for gravitational collapse of a minimally coupled
scalar field in $n$ dimensions. The action of the system is
\begin{eqnarray}\label{S}
S = \int_M d^n x \sqrt{-g} \left[\frac{1}{2\kappa}R -\frac12
(\nabla \phi)^2 \right]+S_{\rm surface},
\end{eqnarray}
where $\kappa\equiv8\pi M_*^{-(n-2)}$ and $M_*$ is the fundamental
scale of the $n$-dimensional gravity theory. The scalar field $\phi$
has a mass dimension of $(n-2)/2$. With the spherical symmetry and the
self-similarity ans\"{a}tz
\begin{eqnarray}\label{sol}
ds^2=e^{2\sigma(\tau)}(-x^2d\tau^2+dx^2)+r^2d\Omega_{n-2}^2,
~~~r=xy(\tau) ~~{\rm and}~~\phi=\phi(\tau),
\end{eqnarray}
after taking into account the surface term the Einstein-Hilbert
action is reduced to
\begin{eqnarray}\label{rs}
S_{\rm red} = V_{n-2}\ell_*^{n-2}\int d\chi d\tau y^{n-4}\left[
\frac{y^2\ddot{\sigma}}{\kappa(n-2)}+\frac{n-3}{2\kappa}\left(-\dot{y}^2
-y^2+e^{2\sigma}\right)+\frac{y^2\dot{\phi}^2}{2(n-2)}\right],
\end{eqnarray}
where the dots denote derivative with respect to $\tau$ and we have
introduced a dimensionless parameter $\chi\equiv x^{n-2}/\ell_*^{n-2}$,
where $\ell_*$ $(\equiv M_*^{-1})$ is the fundamental length.
We have integrated out the angular variables, and $V_{n-2}$ is the
volume of $(n-2)$-dimensional unit sphere, i.e.,
$V_{n-2}\equiv\int d\Omega_{n-2}=2\pi^{(n-1)/2}/\Gamma[(n-1)/2]$.
From this reduced action, one can easily see that
$\sigma={\rm constant}$, and after appropriate rescaling of
coordinates $\sigma$ can be set to zero.
The classical equations of motion for $\phi$ and $y$ are then given by
\begin{eqnarray}
\dot{\phi} &=& \sqrt{\frac{(n-2)(n-3)c}{\kappa}}~
\frac{1}{y^{(n-2)}}, \label{eomphi}\\
\dot{y}^2 &=& y^2 -1 + \frac{c}{y^{2(n-3)}}. \label{eomy}
\end{eqnarray}
Here, $c$ ($>0$) is a dimensionless integration constant
and characterizes the initial distribution of the massless
scalar field. It physically represents the gravitational interaction
strength in the collapse and its inverse gives the measure of the
inhomogeneity of the collapse.\footnote{The constant $c$ does the same
role with parameters $\lambda$ and $\zeta_0$ of Refs. \cite{Ghosh} and
\cite{GB}, respectively, in which self-similarly collapsing null fluid
and dust are examined in different spacetime dimensions. There, $\lambda$
and $\zeta_0$ characterize the initial distributions of null fluid and
dust in each collapses. Their inverses also give the measure of the
inhomogeneity of the collapse. For large values of their inverses
one has highly inhomogeneous collapses where the outer shells collapse
much later than the central ones \cite{lemos}. The common characters
of the parameters $c$, $\lambda$ and $\zeta_0$ can be easily observed
from comparing the energy densities of the systems. In fact, we could
choose other parameters that prescribe the initial data, instead of $c$.
However, with the above physical meaning it seems to be more favorable
to choose $c$ than other parameters.}

From Eq. (\ref{eomy}), we see that the collapsing system can be replaced
by the system of a particle (living in $y$ space) with zero energy under
the influence of the potential
\begin{eqnarray} \label{V1}
V_{\rm cl}(y) = \frac12(1- y^2) -\frac{c}{2} y^{-2(n-3)}.
\end{eqnarray}
The collapse is then parallel to the motion of a particle which starts
from infinity ($y=\infty$) and rolls toward zero ($y=0$).

To see when a black hole is formed, we introduce the local mass function
$M(\tau,x)$ such as
\begin{equation}\label{lmf}
1-\frac{2\kappa}{(n-2)V_{n-2}}\frac{M(\tau,x)}{r^{n-3}}=(\nabla r)^2.
\end{equation}
Clearly, if $n=4$, this reduces to the one usually used in
four-dimensional spacetime, and if the spacetime is static, it
gives the correct mass of Schwarzschild black holes in $n$-dimensions
\cite{Myers}.
We can then localize the apparent horizons given by $(\nabla r)^2=0$,
which translates to $\dot{y}^2-y^2=0$. With this fact, from Eq.(\ref{eomy})
we see that the black hole is formed if the value of $y$ reaches
\begin{equation}\label{yah}
y_{\rm AH}=c^{1/2(n-3)}.
\end{equation}

There are three types of solutions depending on the value of the
constant $c$, i.e., subcritical ($c<c^*$), critical ($c=c^*$)
and supercritical ($c>c^*$) solutions, where $c^*$ is a critical
value of $c$ characterized by potential having second order zero,
i.e., $V_{\rm cl}=V'_{\rm cl}=0$ at $y^*=\sqrt{(n-3)/(n-2)}$,
and it is given by
\begin{eqnarray} \label{ccrit}
c_n^* = \frac{1}{(n-3)}\left[\frac{n-3}{n-2}\right]^{n-2} .
\end{eqnarray}
In the supercritical case where $V_{\rm cl}<0$ everywhere, the value
of $y$ reaches $y_{\rm AH}$ and a black hole is formed by the collapse.
On the other hand, in the subcritical case where there exists a region
with $V_{\rm cl}>0$, the value of $y$ turns around at the turning point
and never reaches $y_{\rm AH}$, which is located in the forbidden region.
In other words, the scalar field collapses, interacts and disperses
leaving behind a flat spacetime without forming a black hole.
In the critical case $y$ takes infinite time to reach $y_{\rm AH}$
and $y=0$, and the spacetime is asymptotically flat but contains a null,
scalar-curvature singularity.

The local mass function of the black hole formed in the supercritical
collapse is given from Eq.(\ref{lmf}) by
\begin{eqnarray}\label{MrAH}
M_{\rm AH} = \frac{(n-2)V_{n-2}}{2\kappa} r_{\rm AH}^{n-3}
  =\frac{(n-2)V_{n-2}}{2\kappa} x^{n-3} y_{\rm AH}^{n-3}
  =\frac{(n-2)c^{1/2}}{2\kappa}V_{n-2} x^{n-3}.
\end{eqnarray}
Clearly, on the apparent horizon the mass becomes unbounded as
$x\to\infty$. The spacetime fails to be asymptotically flat for
supercritical case. As a result, the total mass of the black hole
cannot be written in power-law form in terms of initial data.
This divergence is an unnatural artifact of self-similarity.
In the self-similar solutions this kind of divergence is unavoidable.
In order to avoid this divergence one has to consider the influx of the
scalar field with a finite duration only. This is equivalent to introducing
a cutoff $x_c$ for the $x$-coordinate. Since this is not allowed in the
strict sense of a self-similar solution, we just suppose that any behavior
of fields in the outer region beyond the boundary $x_c$ does not
seriously disturb the self-similar evolution described by the action
(\ref{rs}). One may then interpret the solution as a nearly self-similarly
collapsing (at least in the inner region of $x<x_c$) solution with a
finite mass of
\begin{eqnarray}\label{fmass}
\frac{M_{\rm BH}}{M_*}=\frac{(n-2)c^{1/2}}{16\pi} V_{n-2}
\left(\frac{x_c}{\ell_*}\right)^{n-3}.
\end{eqnarray}
On the other hand, in Ref.\cite{wang} a more realistic model that
represents such type of collapse in four spacetime dimension is
constructed by a cut and paste method: cutting the spacetime along
a null hypersurface and then joining it to an exterior solution
(e.g., an outgoing Vaidya solution). The mass of the black hole is
finite and always takes the form $M_{\rm BH}\propto (c-c^*)^{1/2}$.
We easily expect from the symmetry of the system that a simple
extension of cut and paste method of Ref\cite{wang} to higher dimensions
leads to a similar result
\begin{eqnarray}\label{cmass}
M_{\rm BH}\propto (c-c^*)^\gamma ,
\end{eqnarray}
where $\gamma\equiv(n-3)/\sqrt{2(n-2)}$ is the scaling exponent and it
was identified in Ref.\cite{soda}.

It is interesting to observe that the critical value $c_n^*$ depends
only on the spacetime dimension $n$, and significantly and monotonically
decreases as the number of dimensions increases as shown in Eq.(\ref{ccrit})
and Figure 1. The value $y_{\rm AH}$ also increases as the number of
dimensions increases. These facts tell that the black hole window for initial
data enlarges as the spacetime dimension increases, because the
parameter $c$ specifies the initial distribution of the massless
scalar field. Let us consider a system collapsing from a given initial
distribution of the scalar field specified by a value $c$ smaller
than the critical value $c_n^*$ in $n$-dimensional spacetime.
A black hole can then be formed through the gravitational collapse
from the initial distribution in higher than $n$ dimensions, while
with the same initial distribution the scalar field will disperse
leaving behind just a flat spacetime without forming a black hole
in $n$ or lower dimensions. Thus, we can conclude that higher
dimensions favor a black hole rather than a flat spacetime
with dispersive scalar field when the massless scalar field
self-similarly collapses.

\begin{figure}[htbp]\label{fig:c1star}
\centerline{
\begin{tabular}{cc}
\epsfig{file=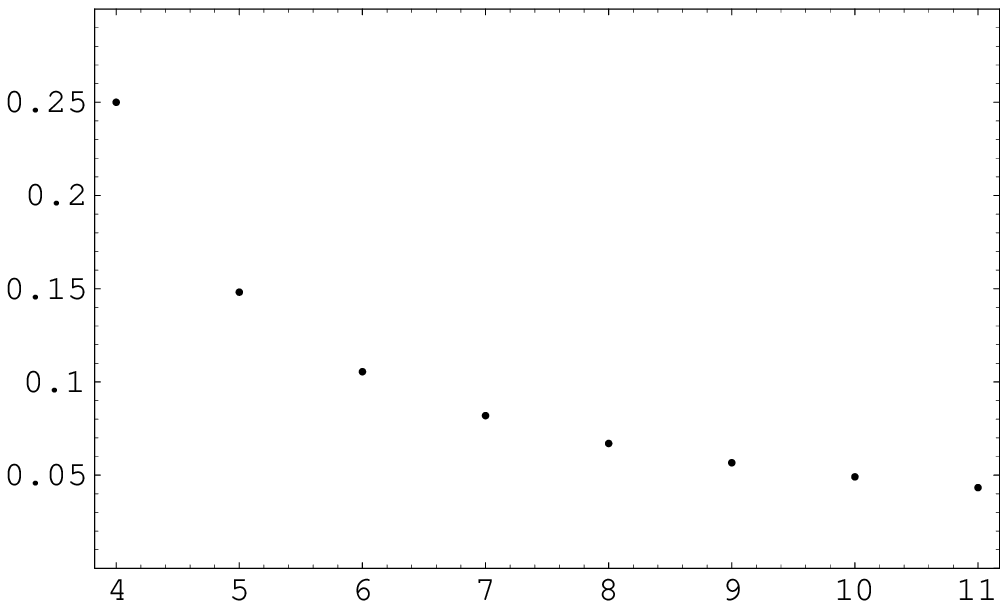, width=7cm} &
\epsfig{file=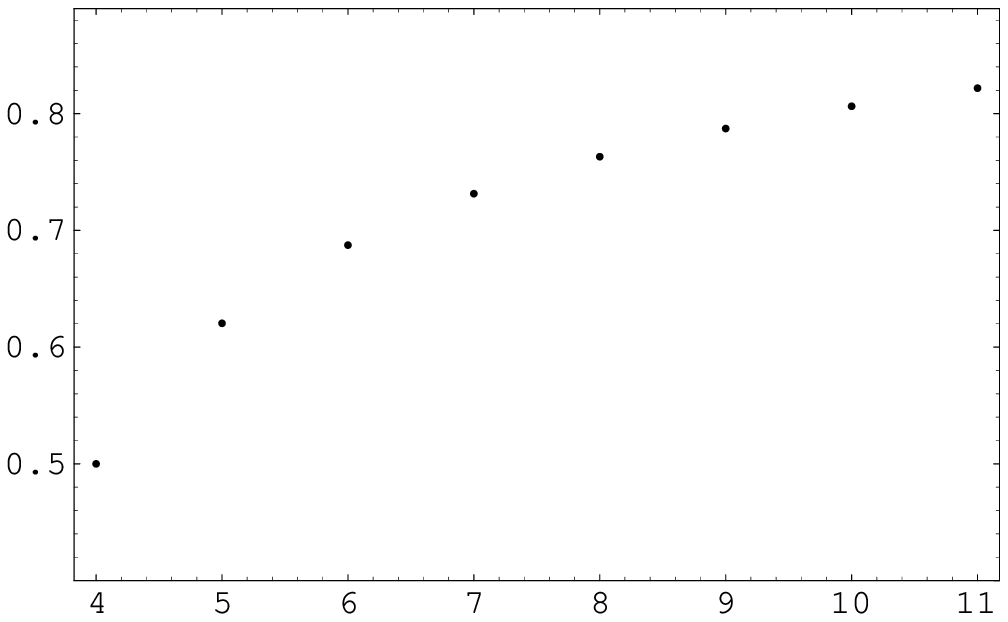, width=6.8cm} \\
$(a)~~~c^*$ & $(b)~~~y_{\rm AH}$
\end{tabular}}
\caption{ Values of $c^*$ and $y_{\rm AH}$ for critical collapse in various
dimensions. }
\end{figure}

Even though we have considered only the self-similarly collapsing scalar
field to study the effect of increase in dimensionality on the results of
gravitational collapse, we would expect similar effects for the dimensional
dependence of black hole formation in different systems from the universality
of the critical phenomena.
In fact, it was already observed in a few different systems.
The authors of Ref. \cite{Ghosh} investigated self-similar
gravitational collapse of null dust fluid in higher-dimensional spacetimes
and showed that as the number of dimensions increases window for naked
singularity shrinks, that is, higher dimensions favor
black hole rather than naked singularity. This could be seen from
the fact that $\lambda_c$ decreases significantly as the dimension
of the spacetime increases, where $\lambda_c$ is the critical value
of the control parameter $\lambda$ which specifies the initial
distribution of the null fluid as the parameter $c$ in our case.
$\lambda_c$ depends on the spacetime dimension $n$ also and has the
same value with $c^*$ as $\lambda_c=(n-3)^{n-3}/(n-2)^{n-2}$.
\footnote{ Here we have redefined $\lambda$ according to our
convention, so this is different by the factor $2^{-(n-2)}$ from
that appears in Ref.\cite{Ghosh}.}
While the final outcome of the collapse is a naked singularity if
$\lambda\leq\lambda_c$, it is a black hole if $\lambda\geq\lambda_c$.
This shows shrinking of the naked singularity window for initial
data. A similar behavior has also been shown in the gravitational
collapse of an inhomogeneous dust cloud described by higher-dimensional
Tolman-Bondi spacetimes in Ref.\cite{GB}, in that the effect of extra
dimensions appears to be a shrinking of the naked singularity initial
data space (of 4D) or an enlargement of the black hole initial data space,
which is characterized by the parameter $\zeta_0$ there.

\section{Quantum collapse in higher dimensions}\label{III}

Black holes are well understood as general-relativistic objects when
their mass $M_{\rm BH}$ far exceeds the higher-dimensional gravity scale
$M_*$. As $M_{\rm BH}$ approaches $M_*$, the black holes will become
quantum gravitational objects and their properties become complex.
This gives rise to an obstacle in calculating the production of black
hole or analyzing the gravitational collapse.
Hence, in the small mass limit it would be physically interesting and
important to study the quantum gravity effect in the collapse.
From the mass formulas Eqs.(\ref{fmass}) and (\ref{cmass}), we see
that there are two independent small mass limits, $x_c\to\ell_*$
and $c\to c^*$. In the case $x_c\sim\ell_*$ the black hole mass is of the
order $M_*$. As $c\to c^*$ it vanishes, so the classical description
at the final stage would not be valid.

It is the purpose of this section to study quantum mechanically the black
hole formation and to investigate how quantum gravity effects modify the
classical picture and the dimensional dependence of gravitational collapse
in higher dimensions. We use the Arnowitt-Deser-Misner (ADM) formulation
to quantize the model and analyze the black hole wave function.
In order to canonically quantize the system, we start from setting
$\sigma=0$ in the metric Eq.(\ref{rs}) and introduce a lapse function
$N(\tau)$ to write the metric as
\begin{equation}\label{metric2}
ds^2=-x^2N^2(\tau)d\tau^2+dx^2+x^2y^2(\tau)d\Omega_{n-2}^2 .
\end{equation}
With this metric the action~(\ref{rs}) is rewritten after taking into
account the boundary term as
\begin{eqnarray}\label{S_y}
S&=& V_{n-2}\ell_*^{n-2} \int d\chi d\tau N y^{n-4}
\left[ \frac{(n-3)}{2\kappa}\left( - \frac{\dot{y}^2}{N^2} -y^2
+1\right)+
\frac{y^2}{2(n-2)}\frac{\dot{\phi}^2}{N^2} \right] .
\end{eqnarray}
The canonical momentum densities of this system for $y$ and $\phi$
are defined as
\begin{eqnarray}\label{p}
\pi_y &=&\frac{\partial{\cal L}}{\partial\dot{y}}
       =-\frac{n-3}{\kappa}\frac{y^{n-4}\dot{y}}{N}, \\
\pi_\phi &=&\frac{\partial{\cal L}}{\partial\dot{\phi}}
          =\frac{1}{(n-2)}\frac{y^{n-2}\dot{\phi}}{N}.
\end{eqnarray}
Then the action $S$ can be rewritten in the ADM formulation as
\begin{eqnarray}
S=V_{n-2}\ell_*^{n-2}\int d\tau d\chi\left[
\pi_y\dot{y}+\pi_\phi\dot{\phi}-N{\cal H}\right],
\end{eqnarray}
where ${\cal H}$ is the effective superspace Hamiltonian density.
The lapse function acts as a Lagrange multiplier, so varying the
action with respect to $N$ we get the Hamiltonian constraint: $H=0$, where
\begin{eqnarray}\label{ham}
H&\equiv&V_{n-2}\ell_*^{n-2}\int d\chi~ {\cal H} \nonumber\\
 &=&V_{n-2}\ell_*^{n-2}\int d\chi\left\{
-\frac{\kappa}{2(n-3)}\frac{\pi_y^2}{y^{n-4}}
+\frac{n-2}{2}\frac{\pi_\phi^2}{y^{n-2}}
-\frac{(n-3)}{2\kappa}(1-y^2)y^{n-4} \right\}.
\end{eqnarray}
Following Dirac's quantization method, the Hamiltonian constraint
becomes a quantum constraint on the wave function
\begin{eqnarray}\label{Psi}
\hat{H}(y,\pi_y;\phi,\pi_\phi)|\Psi> = 0.
\end{eqnarray}
The Hamiltonian constraint is then reduced to
\begin{eqnarray}\label{hc}
\left[-\frac{\kappa}{2(n-3)K}\frac{\hat{\Pi}_y^2}{\hat{y}^{n-4}}
+\frac{n-2}{2K}\frac{\hat{\Pi}_\phi^2}{\hat{y}^{n-2}}
-\frac{(n-3)K}{2\kappa}(1-\hat{y}^2)\hat{y}^{n-4}\right]|\Psi>=0,
\end{eqnarray}
where we have defined the canonical momenta $\hat{\Pi}_i\equiv
V_{n-2}\ell_*^{n-2}\int d\chi\hat{\pi_i}=K\hat{\pi_i}$, and
$K\equiv V_{n-2}\ell_*^{n-2}\int d\chi=V_{n-2}x^{n-2}$.
$K$ represents the volume of $(n-2)$-dimensional sphere with radius $x$.
It has an infinite volume in the limit $x\to\infty$, and the
canonical momenta and the Hamiltonian are divergent. This divergence
is an unnatural artifact of self-similarity as the black hole mass
also diverges. In order to avoid this divergent Hamiltonian we have
to consider the influx of scalar field with finite duration only, as
done to avoid the divergent black hole mass in the previous section,
and we introduce a cutoff $x_c$ in the $x$-coordinate. This cutoff
is related to a scale of the inner region where the assumption of
self-similarity is valid. Since the large $x_c$ limit is the large
black hole limit, it will be equivalent to the classical limit.

When we quantize the system, the first term of equation~(\ref{Psi})
contains an ordering ambiguity between operators $\hat{y}$ and
$\hat{\Pi}_y$. For many choices of the factor ordering, the effect
of the factor ordering can be parameterized by a constant $a$, and
the corresponding Hamiltonian is obtained by the substitution of
$\Pi_y^2\to-y^{-a}[\partial_yy^a\partial_y]$. However, in this paper we
do not adhere to the generality of the operator ordering, and for
simplicity we take the simplest ordering, i.e., $a=0$. Of cause, due
to this-like free fixing of the operator ordering the validity of our
result may be limited. However, we expect that the factor ordering
does not affect any of the semiclassical calculations done in this
paper.\footnote{
For arbitrary $a$, taking $\partial_Y\equiv y^a\partial_y$
the WDW equation (\ref{WDy}) is written as
$[-(1/2)\partial_Y^2+\bar{V}_{WDW}]\bar{\Phi}(Y)=0$,
where $\bar{V}_{WDW}(Y)\equiv y^{2a}V_{WDW}(y)$. On the other hand,
as we will see soon after in the subsection 3. A, the black hole
formation rate depends on the factor
$\alpha\equiv\int dy\sqrt{|V_{WDW}(y)|}$. If we would calculate it with
the semiclassical solution $\bar{\Phi}(Y)$ instead of $\Phi(y)$, then
the rate would depend on
$\bar{\alpha}\equiv\int dY\sqrt{|\bar{V}_{WDW}(Y)|}$. However, we can
easily see that $\alpha=\bar{\alpha}$. Therefore, the factor ordering
does not affect any of the semiclassical calculations.}
We will let the choice of $a$ be dictated by convenience.
The Wheeler-De Witt (WDW) equation~(\ref{Psi}) then becomes
\begin{eqnarray}\label{WDEq}
&&\left[ - \left( \frac{\partial}{\partial y}\right)^2 +
\frac{(n-2)(n-3)}{\kappa y^2}\left(\frac{\partial}{\partial\phi}\right)^2
+\frac{(n-3)^2K_c^2}{\kappa^2}(1-y^2)y^{2(n-4)}\right]
\Psi(y,\phi) =0. \nonumber
\end{eqnarray}
where $K_c\equiv V_{n-2}x_c^{n-2}$.
The wave function can be separated into the scalar field and gravitational
field parts as
\begin{eqnarray}\label{SepVar}
\Psi(\phi, y)= e^{\pm i \Pi_\phi\cdot \phi} \Phi(y) .
\end{eqnarray}
Then the Wheeler-De Witt equation reduces to one-dimensional
Schr\"{o}dinger type equation
\begin{eqnarray}\label{WDy}
\left[-\frac12\left(\frac{\partial}{\partial y}\right)^2+V_{WDW}(y)
\right] \Phi(y) =0,
\end{eqnarray}
where the quantum mechanical potential $V_{WDW}(y)$ is given by
\begin{eqnarray}\label{Vy}
V_{WDW}(y)=\frac{(n-3)^2K_c^2}{2\kappa^2}~y^{2(n-4)}
\left[1-y^2-c\left(\frac{1}{y}\right)^{2(n-3)} \right],
\end{eqnarray}
with $c$ related to the conjugate momentum density $\pi_\phi$ by
$c\equiv [(n-2)/(n-3)]\kappa\pi_\phi^2$. As easily observed from
Eqs. (\ref{eomphi}) and (\ref{p}) $c$ is the same parameter with that
used in the previous section.
This potential is different from its classical counter part (\ref{Vy})
by the multiplication factor $[(n-3)^2K_c^2/\kappa^2]y^{2(n-4)}$.
The shape of the potential in each dimension depends on two parameters,
the initial data $c$ and the cutoff $x_c$. As in the classical collapse,
$c$ determines the supercritical ($c>c^*$), the critical ($c=c^*$)
and the subcritical ($0<c<c^*$) collapses. The cutoff $x_c$ makes the top
of the potential shift away from zero.

In the classical collapse, the collapsing system was replaced by the
system of a particle living in $y$ space with zero energy under the
influence of the potential $V_{cl}$, and the collapse parallels to the
motion of a particle which starts from the infinity ($y=\infty$) and
rolls toward zero ($y=0$). On the other hand, the Wheeler-De Witt
equation (\ref{WDy}) looks like a Schr\"{o}dinger equation, and wave
functions which satisfy the WDW equations describe a particle living in
$y$ space under the influence of the potential $V_{WDW}$. Therefore,
from experiences in quantum mechanics it seems to be plausible to replace
the quantum collapse with the scattering problem of a wave function living
in $y$ space. The boundary condition for the wave functions for black hole
formation then is such that the wave function should be incident from
the spatial infinity in $y$ space, i.e., $y=\infty$, and some part of
them be reflected by the potential barrier back to the infinity of $y$
but the remaining part be transmitted toward the black hole singularity
inside apparent horizon. With armed these boundary conditions, we
compute the rate for black hole formation.

Unfortunately, it is impossible to find exact solutions of the Wheeler-De
Witt equation (\ref{WDy}) except for $n=4$. Therefore, we will try to
see some aspects of quantum collapse by using the WKB method in several
limiting cases: {\it i}) $x_c\gg\ell_*$ and $c\ll c^*$, {\it ii})
$c\sim c^*$. Here, $x_c\gg\ell_*$ means that $x_c$ is large enough for
the top of the potential to be far from zero, so that the WKB is valid
near the top of the potential.

~\\
{\bf A. Subcritical collapse: $x_c\gg\ell_*$ and $c\ll c^*$}\\
~\\
In the case of subcritical collapse, the potential $V_{WDW}$ in any dimension
has two zeros which we will call $y_{n1}$ and $y_{n2}~ (>y_{n1})$.
We then define three separated regions, I ($0<y<y_{n1}$), II
($y_{n1}<y<y_{n2}$) and III~($y>y_{n2}$). In this regime we assume $c$ is
small enough compared to $c^*$ and $x_c$ large enough, so that the
semi-classical analysis valid.
Since we are interested in the collapse of the scalar field starting
from $y=\infty$, we assume that the wave function has a purely incoming
flux toward the singularity at $y=0$ in region I, and the current of
incident wave function in region III is unity for convenience.
With these boundary conditions, the WKB wave functions in each region are
given by
\begin{eqnarray}\label{WKBwf}
\Phi_{\rm I}(y) &=&\frac{i}{\sqrt{p}\cosh\alpha } \exp \left[i
\int_y ^{y_1} p d y - \frac{\pi}{4} i \right], \label{PhiI}\\
\Phi_{\rm II}(y) &=&\frac{i}{2 \sqrt{p} \cosh\alpha }\left[ \exp
\left(- \int_{y_1}^y p d y\right)-i\exp \left(\int_{y_1}^y p d
y\right) \right], \label{PhiII}\\
\Phi_{\rm III}(y) &=&\frac{1}{\sqrt{p}}\exp \left(-i \int_{y_2}^y p d
y- \frac{\pi}{4}i \right)+ \frac{\tanh\alpha}{\sqrt{p}}
 \exp \left(i\int_{y_2}^y p d y+ \frac{\pi}{4} i \right), \label{PhiIII}
\end{eqnarray}
where $p\equiv\sqrt{2|V_{WDW}|}$ and
\begin{eqnarray}\label{alpha}
\alpha\equiv\int_{y_1}^{y_2} p d y = \frac{(n-3)K_c}{\kappa}
\int_{y_1}^{y_2}dy~y^{(n-4)}\left|1-y^2-cy^{-2(n-3)}\right|^{1/2}.
\end{eqnarray}
Except for the phase factor, the wave function (\ref{WKBwf}) depends
only on $\alpha$ which determines the tunnelling and reflection
amplitudes of the waves. The wave function $\Phi_{\rm III}$ indeed satisfies
the boundary condition for black hole formation, in that the
first term describes the incoming component and the second term
describes the outgoing (reflected) component at the far region ($y\gg1$).

The existence of nonvanishing wave $\Phi_{\rm I}$ in region I implies
that there is a possibility to form a black hole quantum mechanically,
even if it is classically forbidden. We now calculate the rate
of a black hole formation in the subcritical collapse.
The incoming flux at spatial infinity computed with Eq.(\ref{PhiIII}) is
\begin{eqnarray}
j_{\rm in}(y) = {\rm Im} \left[\Phi_{\rm in}^*(y)i \hat{\Pi}_y
\Phi_{\rm in}(y)\right]\to 1.
\end{eqnarray}
Some part of the incoming flux is reflected by the potential barrier
back to the infinity $y=\infty$ but the remaining part transmits
toward the black hole singularity at $y=0$. On the other hand, the
reflection rate at the infinity is similarly found as
\begin{eqnarray}
P_{\rm R} =\frac{j_{\rm re}}{j_{\rm in}}= \tanh^2\alpha.
\end{eqnarray}
The rate for black hole formation is the ratio of the transmitted flux
to the incident flux. From the flux conservation we obtain the rate
for black hole formation
\begin{eqnarray}
P_{\rm T}=\frac{j_{\rm tr}}{j_{\rm in}}=1-\frac{j_{\rm re}}{j_{\rm in}}=
{\rm sech}^2\alpha.
\end{eqnarray}
This transmission rate coincides with the rate near $y=0$ obtained by
a direct computation using the transmitted wave Eq.(\ref{PhiI}).

As expected previously, the classical limit is obtained in the large
limit of $K_c$ (or equivalently $x_c$). In a spacetime of dimension $n$,
$\alpha$ depends only on the cut off as $\propto(x_c/\ell_*)^{n-2}$.
As $\alpha$ increases the transmission rate $P_{\rm T}$ exponentially
decreases and vanishes, while the reflection rate $P_{\rm R}$ approaches
to unity. Therefore, most part of the scalar field bounces back without
forming a black hole in the limit of large $x_c$.

\begin{figure}[htbp]\label{pbC}
\centerline{ \epsfig{figure=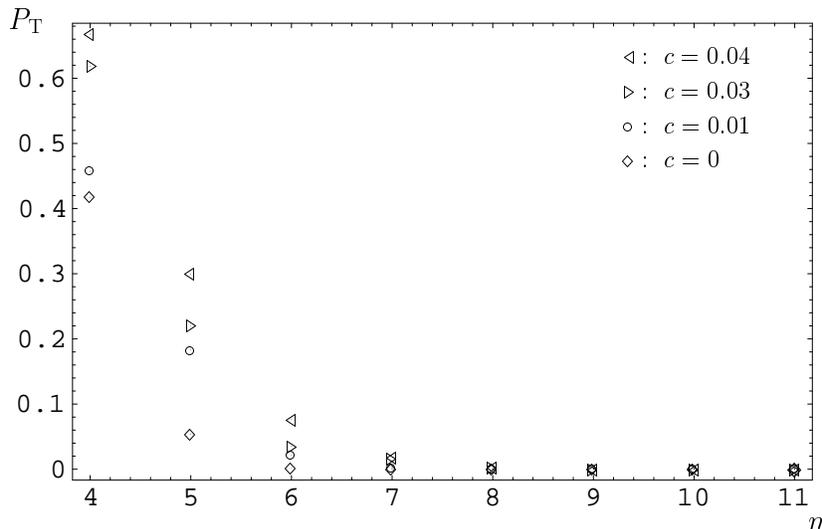,height=7cm}}
\caption{ The rate for black hole formation in diverse dimensions
for various values of the control parameter $c$ and a value of the
cutoff $x_c=1.6 \ell_{Pl} $.}
\end{figure}

On the other hand, the value of $\alpha$ increases as $c$ decreases,
and has a maximum value when $c =0$ for a given cutoff $x_c$ and dimension $n$. The rate for black hole formation falls off as
$c$ decreases, and closes to the minimum rate in the limit $c\to0$:
\begin{eqnarray}\label{ptmin}
P_{\rm T,min}\to{\rm sech}^2\left[ \frac{K_c}{\kappa}\frac{\sqrt{\pi}
\Gamma\left(\frac{n-1}{2}\right)}{2\Gamma\left(\frac{n}{2}\right)}\right]
~~~{\rm as}~~~c\to0 .
\end{eqnarray}
Since the parameter $c$ characterizes the gravitational interaction
strength in the collapse and small $c$ means small interaction
strength, it is reasonable that the rate for black hole formation falls
off as $c$ becomes small. However, it seems surprising that there exists
a minimum rate for black hole formation with vanishingly small
gravitational interaction strength of limit $c\to0$. This is not an
artifact of our approximation by WKB method. We can also observe this
aspect in Ref. \cite{BKKSY}, in which the transmission rate is exactly
calculated using the analytic solution of Eq. (\ref{WDy}). Unless $c=0$
definitely, the potential is singular at the origin at $y=0$, and the
incident waves are not reflected at the origin.\footnote{The case
$c=0$ should be distinguished from the limit $c\to0$. If $c=0$ definitely,
the potential is not singular but is zero at $y=0$. The incident waves
are totally reflected at the origin, so the transmission rate is then
zero. Notice that the spacetime described by the metric Eq. (\ref{sol})
does not have a singularity, but is regular at the origin.}
Note that the curvature singularity at the origin acts like a sink for
the incident waves. Therefore, the transmission rate does not become
zero in the limit $c\to0$. The minimum rate may have a different form
from Eq. (\ref{ptmin}), because in the limit $c\to0$ the potential
becomes very flat near $y=y_{n1}$ and is very close to zero in a broad
range. So our WKB approximation would not be valid near $y=y_{n1}$.
The numerical works in Fig. \ref{pbC} confirm the nonvanishing
transmission rate in the limit $c\to0$.
On the other hand, we have to consider the position of the apparent horizon.
If the apparent horizon is at $y=0$ as in the classical collapse, then there
is no black hole formation despite of the existence of the transmission
flux in the region I. However, the apparent horizon and black hole mass
will be affected by quantum effect. The quantum mechanical effect will shift
the position of the apparent horizon from $y=0$ to $y\sim 1/K_c^2$ as
discussed in Ref.\cite{BKKSY}.


The rate for black hole formation doesn't seem to vary simply with
spacetime dimensions for a fixed initial data $c$ and cutoff $x_c$.
The $n$ dependence of $\alpha$ is complicated as can be seen from
Eq.(\ref{alpha}).
The volume of $(n-2)$-dimensional unit sphere $V_{n-2}$ included in $K_c$
does not monotonically behave but has a maximum value at $n=7$, while
the other factors monotonically increase as $n$ increases.
However, with some generic values of the cutoff $x_c$, $\alpha$ behaves
simply and $P_{\rm T}$ monotonically decreases within four to eleven
dimensions, in which we are interested from
the unified theory perspective like string theory.
Since the integral in Eq.(\ref{alpha}) is not explicitly calculable in
general dimensions, we present numerical values of $P_{\rm T}$ for several
different $c$ in Figure 2.
The result shows that the rate for black hole formation by quantum
tunneling falls off as the number of dimensions increases for a given
initial data $c$ and fixed cutoff $x_c$.

~\\
{\bf B. Near-critical collapse: $c\sim c^*$}\\
~\\
In this regime we cannot directly apply the usual WKB approximation,
because the height of the potential is too low and the derivative of
the potential is almost zero near turning points. So the semiclassical
solution such as $\Phi_{\rm II}$ of previous case is not valid near
the top of the potential. Note that WKB is valid only when the condition
$\delta\lambda/\lambda\ll1$ is satisfied, where $\lambda$ is the wave
length of the wave functions. The spirit of the WKB method is that one
find a solution of the wave equation near the turning point by
approximating the potential with a linear function. Then one matches
the solution with the other two semiclassical solutions outside the
transition region.

We develop an approximation which has a common spirit with the usual
WKB method. We approximate the potential $V_{WDW}$ with a quadratic
function near the position of the top and find an analytic solution
of the wave equation with the quadratic potential. We then match the
solution with the other two semiclassical solutions in regions I and
III choosing appropriate coefficients. This approximation method is
not so different from the conventional WKB method but only a slight
deformation from the WKB method, besides that we introduce a quadratic
approximation instead of a linear approximation. So it has much common
sprit with the usual WKB method, and it will give approximately good
answer as much as the usual WKB method does.

We represent the near critical limit by setting $c=c^*(1+\delta)$ with
$|\delta|\ll1$, where the collapse is supercritical if $\delta>0$ and
subcritical if $\delta<0$. The Wheeler-De Witt potential is then
separated as
\begin{eqnarray}\label{svwdw}
V_{NC}(y)=V_{\rm crit}(y)-\frac{(n-3)^2K_c^2}{2\kappa^2y^2}c^*\delta,
\end{eqnarray}
where $V_{\rm crit}(y)$ is potential for critical collapse and can be
written as
\begin{eqnarray}\label{Vcrit}
V_{\rm crit}(y)=-\frac{(n-3)^2K_c^2}{2\kappa^2}\left(y^2-{y^*}^2\right)^2~
\sum_{k=0}^{n-4}\frac{k+1}{n-2}~{y^*}^{2(n-k-5)}y^{2k-2}.
\end{eqnarray}
$V_{\rm crit}(y)$ has a second order zero at $y^*=\sqrt{(n-3)/(n-2)}$ as
the classical potential $V_{\rm cl}$ does. Near $y=y^*$, the potential
may be approximated as
\begin{eqnarray}\label{VNC}
V_{NC}&\simeq&-\frac{(n-3)^2K_c^2}{2\kappa^2y^2}
\left[c^*\delta+\frac{n-3}{2} {y^*}^{2(n-5)}~(y^2-{y^*}^2)^2\right] \nonumber\\
&=& E - \frac{V_+}{2}y^2 - \frac{V_-}{2y^2},
\end{eqnarray}
where
\begin{eqnarray}
V_- &=&\frac{(n-3)^3K_c^2}{2\kappa^2}~{y^*}^{2(n-3)}
       \left[1+\frac{2c^*\delta}{n-3}~{y^*}^{-2(n-3)}\right] ,\\
V_+ &=&\frac{(n-3)^3K_c^2}{2\kappa^2}~{y^*}^{2(n-5)}, \\
E &=&\frac{(n-3)^3K_c^2}{2\kappa^2}~{y^*}^{2(n-4)} .
\end{eqnarray}
In fact, there could be several other choices in choosing the approximate
quadratic potential. But we have chosen above approximate potential upon
the following advantages: First, with this choice the WDW equation is
exactly solvable. Second, it well approximates the full potential near
the origin at $y=0$ as well as around the top, so the solutions for the
approximate potential behaves like $\Phi_I$ in the region I.
Lastly, in 4-dimensional case, it corresponds to the exact potential,
not approximate one.

With above approximate potential, the Wheeler-De Witt equation
allows solutions in terms of the confluent hypergeometric function \cite{AS}:
\begin{eqnarray}\label{NCwf}
\Phi_{\rm NC}(y)=De^{-i\sqrt{V_+} y^2/2}\left(i\sqrt{V_+}
y^2\right)^{\mu} M(a,b,i \sqrt{V_+} y^2),
\end{eqnarray}
where $D$ is a constant and
\begin{eqnarray}
\mu &\equiv& \frac{1}{4} - \frac{i}{2} Q, \\
a &\equiv& \frac{1}{2} -\frac{i}{2} (Q+\sqrt{E}y^*),\nonumber \\
b &\equiv& 1- i Q,  \nonumber
\end{eqnarray}
with $Q = (Ey^{*2} \nu^2 - 1/4)^{1/2}$, $\nu= V_-/(Ey^{*2})$. There exists
another independent solution of the Wheeler-De Witt equation with the
potential (\ref{VNC}). However, we will not write down that solution,
because the solution (\ref{NCwf}) is enough to describe the formation of
a black hole. Remind that we follow the analogy to the scattering problem
of quantum mechanics and the boundary conditions mentioned previously.
According to the boundary conditions, it seems reasonable to choose only
purely ingoing wave functions in region I. The same wave function was
used to describe the four-dimensional formation of a black hole by a
scalar field collapse~\cite{BKKSY}.

We now match this solution with the semiclassical solutions in regions
I and III. In fact, the solution (\ref{NCwf}) is also valid near $y=0$,
because the approximate potential $V_{NC}$ has the same asymptotic form
as the full potential $V_{WDW}$. Therefore, we only need to match the
solution (\ref{PsiNC}) with the semiclassical potential in region III.
Using the asymptotic form of the confluent hypergeometric function
\begin{eqnarray}\label{Psi(infty)}
\lim_{|z|\rightarrow \infty}M(a,b,z) \approx\Gamma(b)
\left[\frac{e^{i \pi a} z^{-a}}{\Gamma(b-a)}+ \frac{e^z
z^{a-b}}{\Gamma(a)}\right], ~~ -\frac{\pi}{2} < {\rm arg}(z) <
\frac{3\pi}{2} ,
\end{eqnarray}
we get the asymptotic form of the wave function at large $y$:
\begin{eqnarray}\label{PsiNC}
\Phi_{\rm NC}(y)
\approx D\left\{\frac{\Gamma(b)}{\Gamma(b-a)}
e^{i\pi a}\left(i\sqrt{V_+}y^2\right)^{\mu-a}e^{-i\sqrt{V_+}y^2/2}
+\frac{\Gamma(b)}{\Gamma(a)}\left(i\sqrt{V_+}y^2\right)^{\mu+a-b}
e^{i\sqrt{V_+}y^2/2}\right\},
\end{eqnarray}
where $a-\mu=1/4-i\sqrt{E}y^*/2$ and $b-a-\mu=1/4+i\sqrt{E}y^*/2$.
The first term describes the incoming wave function and the second term
describes the outgoing wave function.
Comparing the asymptotic form (\ref{PsiNC}) with the semiclassical
potential in region III of the form
\begin{eqnarray}\label{WKB}
\Psi(y) =\frac{ D_1}{\sqrt{p(y)}} e^{i \int^y p(y') dy'}+\frac{
D_2}{\sqrt{p(y)}} e^{-i \int^y p(y') dy'},
\end{eqnarray}
we obtain the asymptotic form of the wave function in the region III
\begin{eqnarray}\label{WKB2}
\Psi_{\rm III}(y)\approx\tilde{D}\left\{\frac{\Gamma(b)e^{i\pi
a}}{\Gamma(b-a)} \left[\frac{\sqrt{V_+}}{p^2(y)}\right]^{a-\mu}
e^{-i\int^y p(y')dy'} +\frac{\Gamma(b)}{\Gamma(a)}\left[\frac{
\sqrt{V_+}}{p^2(y)}\right]^{b-a-\mu} e^{i \int^y p(y')
dy'}\right\},
\end{eqnarray}
where $p= \sqrt{2 V_{WDW}}$.
Note that the asymptotic solution (\ref{PsiNC}) is of the same form as
the semiclassical solution (\ref{WKB}) with $\bar{p}\equiv\sqrt{-2V_{NC}}$.

Then the rate for black hole formation is easily obtained from the two
solutions (\ref{PsiNC}) and (\ref{WKB2}):
\begin{eqnarray}\label{PTNC}
P_{\rm T} = \frac{j_{\rm tr}}{j_{\rm in}} =
\frac{e^{-\frac{\pi}{2}(\sqrt{E}y^*+Q)} \sinh \pi Q}{\cosh
\frac{\pi}{2}(\sqrt{E}y^*-Q) }.
\end{eqnarray}
It is easy to check that this transmission rate coincides with that obtained
by using current conservation at the spatial infinity
\begin{eqnarray}
P_{\rm R}= \frac{j_{\rm re}}{j_{\rm in}}
=\frac{\cosh\frac{\pi}{2}(Q+\sqrt{E}y^*)}{\cosh\frac{\pi}{2}(Q-\sqrt{E}y^*)}
e^{-\pi Q} = 1- \frac{j_{\rm tr}}{j_{\rm in}}.
\end{eqnarray}
This determines the probability for black hole formation in the near critical
collapse for values of $|\delta|\ll1$.

In the large cutoff limit $x_c\gg\ell_*$ it will be easier to see the
behavior of the rate $P_{\rm T}$.
In the large $x_c$ limit the rate for black hole formation should be
calculated separately for $\delta>0$ and $\delta<0$.
For $\delta>0$, Eq.(\ref{PTNC}) becomes asymptotically
\begin{eqnarray}\label{ptn1}
P_{\rm T}\approx 1-\exp\left[-\frac{\sqrt{2} \delta K_c}{\sqrt{n-3} \kappa}
\left(\frac{n-3}{n-2}\right)^{\frac{n}{2}}\right].
\end{eqnarray}
In the case $\delta<0$, the asymptotic value of Eq.(\ref{PTNC}) in the large
$x_c$ limit is
\begin{eqnarray}\label{ptn2}
P_{\rm T}\approx \exp\left[\frac{\sqrt{2} \delta K_c}{\sqrt{n-3} \kappa}
\left(\frac{n-3}{n-2}\right)^{n/2}\right].
\end{eqnarray}
This is the tunneling rate for black hole formation in the subcritical
collapse near the critical case. The case that $\delta=0$ has to be considered
separately. The asymptotic form is then
\begin{eqnarray}\label{ptn3}
P_{\rm T}=\frac{e^{-\frac{\pi}{2}(\sqrt{V_-}+\sqrt{V_--1/4})}
\sinh\pi\sqrt{V_--1/4}}{\cosh\frac{\pi}{2}(\sqrt{V_-}+\sqrt{V_--1/4})}.
\end{eqnarray}
As in the previous case, the classical limit is obtained in the very large
cutoff limit $x_c\gg\ell_*$. The rate $P_{\rm T}$ exponentially depends
on $K_c$. In the very large $x_c$ limit $P_{\rm T}$ becomes unity for
$\delta>0$, which implies that a black hole is formed from the collapsing
scalar field as in the classical case.
For $\delta<0$, $P_{\rm T}$ becomes zero. This tells that the scalar field
bounces back without forming a black hole.
For $\delta=0$, $P_{\rm T}$ becomes $1/2$.

\begin{figure}[htbp]\label{pb}
\centerline{ \epsfig{figure=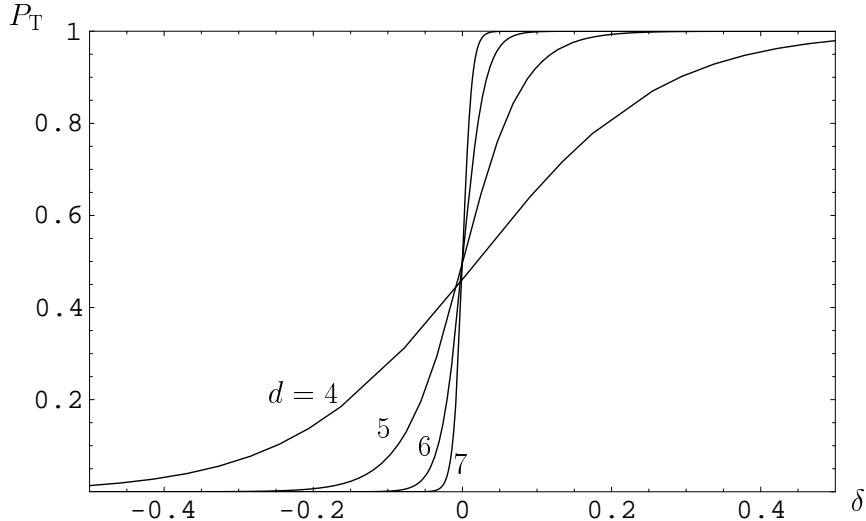,height=7cm}}
\caption{The rates for black hole formations in near critical collapses
in diverse dimensions for a value of the cutoff $x_c=1.5 \ell_{Pl} $.}
\end{figure}

On the other hand, the dimensional dependence of the rate $P_{\rm T}$ is
more complicated, because the volume factor $V_{n-2}$ included in $K_c$
does not behave simply but reaches to the maximum value at $n=7$, while
the product of all the other factors in the exponent of Eq.(\ref{ptn1})
monotonically increases with fixed $c$ and $x_c$ as dimension $n$
increases. However, for generic values of $x_c$ the absolute value of the
exponent monotonically increases between four to eleven dimensions.
As the spacetime dimension $n$ increases the rate for black hole
formation $P_{\rm T}$ increases for $\delta>0$, while $P_{\rm T}$ decreases
for $\delta<0$. Therefore, the rate for black hole formation by quantum
tunnelling falls off as dimension increases in the subcritical collapse
with $\delta<0$, while black hole formation is easier in higher dimensions
for $\delta>0$, as shown in Fig. \ref{pb}. This implies that the aspect of
quantum collapse becomes close to the classical aspect as dimension
increases. This can be interpreted as the quantum gravity effects on
the result of gravitational collapse lessens with increase of the
spacetime dimensionality.

\section{Discussions: gravitational collapse in a brane world with large
but compact extra dimensions}

We have considered self-similarly collapsing system of a massless scalar
field in higher dimensions, both classically and quantum mechanically.
We examined the dimensional dependence of such gravitational collapse
and formation of black hole. Higher dimensional spacetime seems to favor
formation of black hole rather than other final outcomes, in that the initial
data space for black hole formation enlarges as dimension increases, i.e.,
the critical value $c_n^*$ of the control parameter which describes initial
distribution of the scalar field falls off as dimension increases.

We have studied black hole formation quantum mechanically by the collapsing
scalar field. We canonically quantized the system and found the wave
functions by using the WKB method. The results of quantum
collapse obtained in four-dimensional spacetime \cite{BKKSY} go over to
higher-dimensional quantum collapse. The wave functions for black hole
formation have not only incoming flux but also outgoing flux in both the
subcritical and supercritical cases. The classical evolution of black hole
formation is recovered in the limit of large cutoff $x_c$. On the other hand,
the rate for black hole formation by quantum tunnelling in the subcritical
collapse lessens as spacetime dimension increases, that is, it becomes harder
to form a black hole in higher dimension in the subcritical collapse.
The reflection rate in the supercritical (but close to the critical)
collapse, however, decreases as dimension increases: The black hole
formation is easier in higher dimensions in the supercritical collapse.
This implies that the quantum gravity effect on the gravitational collapse
lessens with the increase of spacetime dimension.

We now discuss the gravitational collapse in a brane world with large
but compact extra dimensions first explored by Arkani-Hamed et. al.
\cite{ADD}. The dimensional dependence of the final result in gravitational
collapse implies that if we are living in the brane world with large extra
dimensions, then the aspect of gravitational collapse is drastically
different from that expected in usual four-dimensional world.
Even though we prepare a macroscopic initial distribution of matter in
four-dimensional spacetime, which final product is resulted from the collapse
will depend on the number of extra dimensions, because if the Schwarzschild
radius is smaller than the size of extra dimensions, the collapse will be
higher-dimensional rather than four-dimensional at the final stage of the
collapse.

Suppose that one prepares an initial distribution of the scalar field, of
which typical size is much larger than the extra dimension size, so that the
collapse is effectively four-dimensional at the early stage of the collapse.
Here, we assume that the three-brane has negligible self-gravity of its own
as in usual large extra dimension scenarios and so the brane can be identified
as a surface of vanishing extrinsic curvature. Therefore, if we restrict
to black holes of size $r_{\rm BH}$ much less than the size of extra
dimensions $L$, then the geometry near the black hole will be very well
approximated by a $n$-dimensional Schwarzschild solution \cite{Myers}.
If a black hole is formed from matters on the brane, the symmetry requires
that the brane pass through the equator of the black hole.
If the initial distribution is prepared (taking $c>c_4^*$ and large enough
$x_c$) to form a black hole with Schwarzschild radius $r_{\rm BH}$ larger
than the extra dimension size $L$ ($r_{\rm BH}\gg L$), then the collapse will
be effectively four-dimensional event to an observer outside the horizon,
because the observer can not see the collapse after the apparent horizon forms.
The geometry of the resulting black hole is simply a product of
four-dimensional Schwarzschild solution and a $(n-4)$-dimensional torus,
and its dynamical properties are determined by four-dimensional physics.

However, if the initial distribution is prepared (if a black hole is formed)
so that the resulting black hole have Schwarzschild radius of
$\ell_*\ll r_{\rm BH}\ll L$, the collapse goes under the size of the extra
dimensions and becomes essentially $n$-dimensional at the final stage of the
collapse.\footnote{The mass of the black hole can still be much larger than
the four-dimensional Planck scale and can be treated as a general relativistic
object. In the usual large extra dimension scenario \cite{ADD},
the size of extra dimensions $L$ is related to the effective four-dimensional
Planck scale $M_{\rm Pl}$ and the $n$-dimensional gravity scale $M_*$ by
$L\sim M_*^{-1}(M_{\rm Pl}/M_*)^{2/(n-4)}$, while the Schwarzschild radius of
$n$-dimensional black hole is $r_{\rm BH}=M_*^{-1}[16\pi
M_{\rm BH}/(n-2)V_{n-2}M_*]^{1/(n-3)}$. So the black hole mass of size
$r_{\rm BH}\sim L$ is $M_{\rm BH}\sim M_*(M_{\rm Pl}/M_*)^{2(n-3)/(n-4)}$.
If $M_*$ is of order a TeV, then $M_{\rm BH}\sim 10^{32(n-3)/(n-4)}M_*
\sim 10^{32(n-3)/(n-4)-21}{\rm g}$. This is much larger than the
four-dimensional Planck mass, e.g., for $n=6$, $M_{\rm BH}\sim 10^{27}{\rm g}$
and, for $n=10$, $M_{\rm BH}\sim 10^{16}{\rm g}$, about the mass of the
Earth and the typical mass of primordial black holes, respectively.}
Therefore, the final result of the collapse can be different from that
expected in the purely four-dimensional world.
If one prepares the initial distribution specified by $c>c^*_4$, then
a black hole is formed as expected in four-dimensional world because
$c_4^*>c_n^*$. Since, for $r\gg L$, the geometry
of the resulting black hole is well approximated by the four-dimensional
Schwarzschild black hole and the mass measured on the brane is the same as
the mass in the bulk, it may look like a usual four-dimensional
Schwarzschild black hole. However, it has different thermodynamic properties
from those of a purely four-dimensional black hole \cite{EHM}.
The Schwarzschild radius is larger than it would be for a four-dimensional
black hole. This means that the temperature is lower, the evaporation rate
is slower, and the horizon area (the entropy of the black hole) is larger.
If $c<c^*_n$, then there is no black hole formation and the scalar field
disperses back to the four-dimensional world. The spectrum of the dispersed
scalar field will not be like that in a purely four-dimensional collapse,
because the very light Kaluza-Klein modes with masses starting at $\sim1/L$
couples to the matter and some fraction of the energy radiate off of the
brane into the bulk.

The case that the initial distribution is prepared to have the control
parameter of $c_n^*<c<c_4^*$ may be more interesting.
Then, an $n$-dimensional black hole will be formed and there will be no
dispersing scalar field. This is not the one expected if we are living
in purely four-dimensional world. If our world is purely
four-dimensional, or the extra dimension is very small and of the order of
quantum gravity scale as taken in usual higher-dimensional unified theories,
then there will be no black hole formation and the scalar field will
disperse back to the four-dimensional world.
Therefore, we can expect that one can probe the existence and
determine the number of large extra dimensions by examining the gravitational
collapse and the dimensional dependence of black hole formation, (at least)
in principle.

Of course, we have over-simplified the problem. At the final stage
when the collapse effectively occurs in $n$ dimensions, the collapse
is not like the $n$-dimensional spherical collapse discussed in the
previous sections. The collapse is not spherically symmetric and
self-similar, even if we have prepared the initial distribution to be
spherically symmetric and the collapse to occur self-similarly at the
beginning, at which the collapse can be described by the four-dimensional
general relativity. The collapse may look like that of a disk-like rather
than a spherical distribution at the late stage, because the matter does
live only on the three-brane. Therefore, the discussions in the previous
two paragraphs suffer from lack of physical meaning. Despite of this poor
physical meaning, the purpose we presented the discussions upon the
over-simplified assumption is simply to illustrate the probing mechanism
of the large extra dimensions. In this sense, the discussions in the two
previous paragraphs seems to have its necessity. For rigorous treat, we
have to consider the collapse of disk-like distribution of matters and
the critical phenomena. However, it seems impossible to find analytic
solutions for collapses of disk-like matter distributions. Some numerical
simulations for such collapses may be helpful. But, this is beyond the
scope of this paper.

In spite of all, we easily expect that the brane world scenario makes
interesting predictions about the formation of small black holes.
In particular, the existence of the large extra dimensions would affect
on the formation of primordial black holes, which are formed as a result
of the density fluctuation in the early universe. If there was large extra
dimensions, then mass spectrum of the primordial black hole would not be
like that predicted in purely four-dimensional theory.
It will be interesting to investigate their detailed phenomenology.

~\\
{\Large {\bf Acknowledgments}} \\
~\\
This work was supported in part by BK21 Project of Ministry of
Education (S.-H.M. and J.H.Y.) and by Korea Research Foundation under
Project number 99-005-D00010 (H.-C.K. and J.H.Y.).



\end{document}